\documentclass[conference,a4paper]{IEEEtran}

\usepackage[utf8]{inputenc} 
\usepackage[T1]{fontenc}
\usepackage{url}              
\usepackage{cite}             

\usepackage[numbers,sort&compress]{natbib}

\usepackage[cmex10]{amsmath}  
\usepackage{mleftright}       
\mleftright                   

\usepackage{graphicx}         
\usepackage{booktabs}         

\usepackage{pgfplots}
\pgfplotsset{compat=newest}
\usepgfplotslibrary{fillbetween}

\usepackage{amsmath}
\usepackage{amssymb}
\usepackage{mathrsfs}
\usepackage{bbm}
\usepackage{comment}
\usepackage{color}
\usepackage[normalem]{ulem}
\newtheorem{theorem}{Theorem}
\newtheorem{lemma}{Lemma}
\newtheorem{definition}{Definition}
\newcommand{\red}[1]{\textcolor{red}{#1}}
\newcommand{\klcomment}[1]{{\bf{{\red{{{KL --- #1}}}}}}}

\allowdisplaybreaks

\newcommand{\aln}[1]{\begin{align*}#1\end{align*}}
\newcommand{\al}[1]{\begin{align}#1\end{align}}

\newcommand{\R}{{\mathbb R}}
\newcommand{\x}{{\bf x}}
\newcommand{\bfv}{{\bf v}}

\newcommand{\Y}{{\mathcal Y}}
\newcommand{\X}{{\mathcal X}}

\newcommand{\A}{{\mathcal A}}
\newcommand{\F}{{\mathcal F}}

\newcommand{\E}{{\mathcal E}}
\newcommand{\B}{{\mathcal B}}
\newcommand{\C}{{\mathcal C}}
\newcommand{\D}{{\mathcal D}}

\newcommand{\cH}{{\mathcal H}}

\renewcommand{\emph}[1]{{\textit{#1}}}

\begin{document} 

\title{Fundamental Limits of Multiple Sequence Reconstruction from Substrings}

%
\author{%
   \IEEEauthorblockN{Kel~Levick}
   \IEEEauthorblockA{University of Illinois, Urbana-Champaign\\
                     Urbana, IL, USA\\
                     klevick2@illinois.edu}
                   \and 
    \IEEEauthorblockN{Ilan~Shomorony}
   \IEEEauthorblockA{University of Illinois, Urbana-Champaign\\
                     Urbana, IL, USA\\
                     ilans@illinois.edu}
            }

\maketitle

\begin{abstract}
The problem of reconstructing a sequence from the set of its length-$k$ substrings has received considerable attention due to its various applications in genomics. 
We study an uncoded version of this problem where multiple random sources are to be simultaneously reconstructed from the union of their $k$-mer sets. We consider an asymptotic regime where $m = n^\alpha$ i.i.d.~source sequences of length $n$ are to be reconstructed from the set of their substrings of length $k=\beta \log n$, and seek to characterize the $(\alpha,\beta)$ pairs for which reconstruction is information-theoretically feasible.
We show that, as $n \to \infty$, the source sequences can be reconstructed if $\beta > \max(2\alpha+1,\alpha+2)$ and cannot be reconstructed if $\beta < \max( 2\alpha+1, \alpha+ \tfrac32)$, characterizing the feasibility region almost completely.
Interestingly, our result shows that there are feasible $(\alpha,\beta)$ pairs where repeats across the source strings abound, and non-trivial reconstruction algorithms are needed to achieve the fundamental limit.

\end{abstract}

\section{Introduction}

The problem of reconstructing a sequence from the set of its length-$k$ substrings (or $k$-mers) has a long history, dating back to the development of the  sequencing-by-hybridization (SBH) technology in the late 1980s \cite{dramanac1989SBH}.
Due to this technological advance, significant attention was originally devoted to understanding when the set of $k$-mers uniquely determines the underlying sequence, and to designing algorithms to perform the reconstruction efficiently \cite{ukkonen,pevzner1989tuple,pevzner2001eulerian,skiena1995reconstructing}.

Inspired by Ukkonen's 1992 paper~\cite{ukkonen}, which characterized how large $k$ needs to be to allow reconstruction, a more recent line of work considered this problem from an information-theoretic perspective~\cite{preparata2000sequencing,motahari_shotgun,mohajer2013reference,motahari_noisy,shomorony_errors}.
In particular, \cite{motahari_shotgun} considered a random  source $\x$ (modeling a genome sequence), and 
provided conditions on how large $k$ needs to be (and how many length-$k$ substrings need to be observed) in order for the reconstruction of a source sequence $\x$ to be information-theoretically feasible.

In this work, we consider a multiple-source version of this problem.
We assume that there are $m$ random source sequences $\x_1,\dots,\x_m$, each of length $n$, and we wish to reconstruct them from the union of their sets of $k$-mers.
This multiple-source setting can be motivated by modern sequencing assays such as metagenomics and RNA-seq, where one observes substrings from a mixture of distinct source sequences (different microbial genomes in the case of metagenomics and different mRNA molecules in the case of RNA-seq).
It can also be motivated by the fact that many bioinformatics algorithms first convert a set of reads into the set of their constituent $k$-mers and then operate in $k$-mer space. It is thus relevant to understand when such sets fully preserve the information in the original reads.
Our setting is also connected with the problem of DNA-based data storage, where one wishes to recover a set of $m$ synthesized DNA molecules by jointly sequencing them.
Our work can thus be seen as an \emph{uncoded} version of the recent work on multi-strand reconstruction from substrings~\cite{multistrand-yehezkeally}, or a random coding approach to it (see also ``Related work'' below). 






\begin{figure}[b!]
    \centering
    \vspace{-8mm}    \includegraphics[width=0.99\columnwidth]{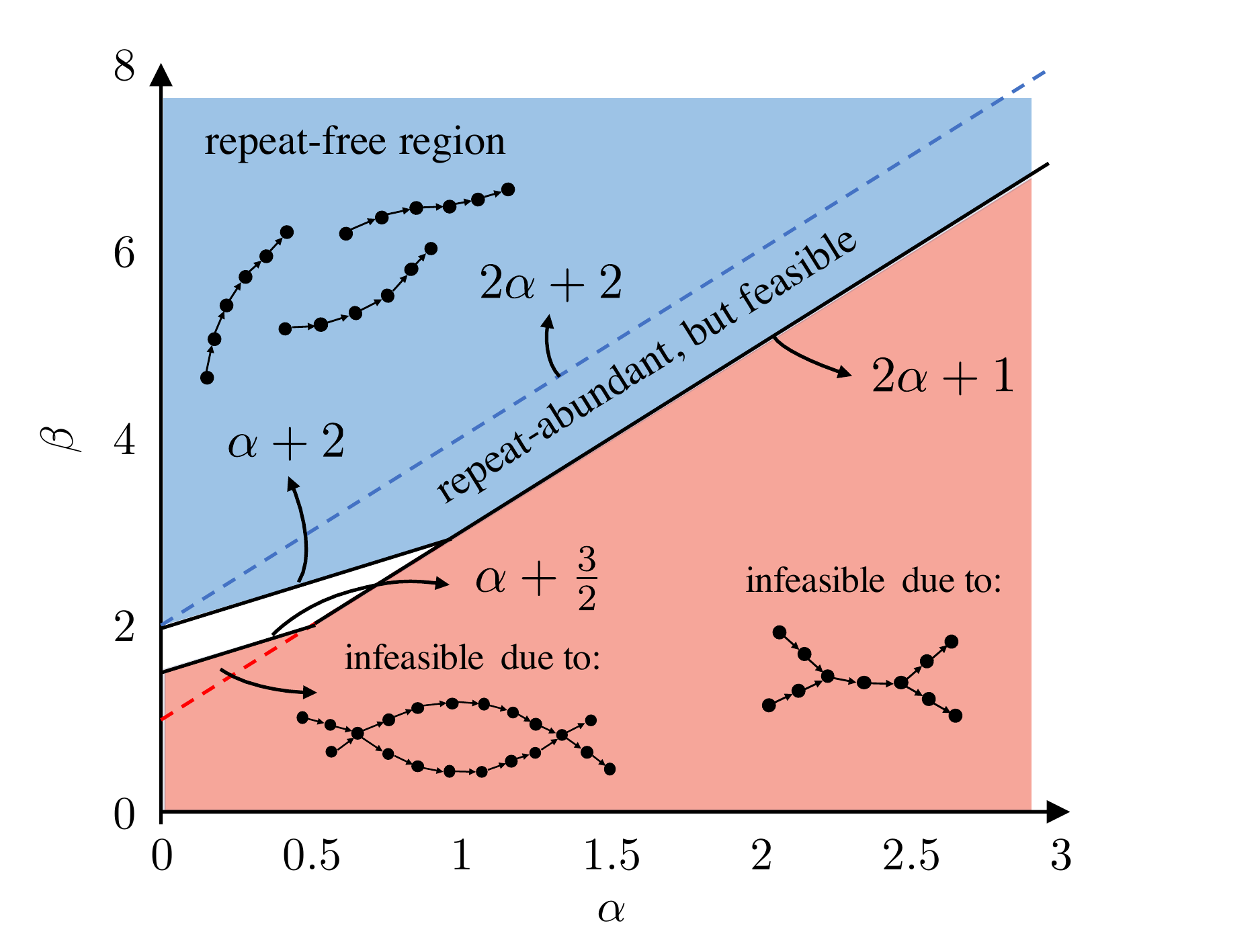}
    \caption{
    The blue region is given by $\beta > \max(2\alpha+1,\alpha+2)$, and corresponds to 
    $(\alpha,\beta)$ pairs for which $\x_1,\dots,\x_m$ are uniquely reconstructible from the set of $k$-mers with vanishing error probability.
    The red region is given by $\beta< \max(2\alpha+1,\alpha+2)$, and corresponds to $(\alpha,\beta)$ pairs where the solution is not unique. 
    We illustrate the types of $k$-mer de Bruijn graphs that lead to the ambiguity.
    The white region corresponds to the unknown $(\alpha,\beta)$ pairs.} 
    \label{fig:region}
\end{figure}

Our goal is to characterize how large $k$ needs to be to allow perfect reconstruction of all $m$ source sequences.
We assume each source is i.i.d.~${\rm Bern}(\tfrac12)$ and consider an asymptotic regime where $m = n^\alpha$ and $k=\beta \log n$, for some constants $\alpha,\beta > 0$.
We aim to characterize for which $(\alpha,\beta)$ pairs the set of $k$-mers uniquely determines the set of source sequences $\{\x_1,\dots,\x_m\}$ with vanishing error probability.
Our main result, illustrated in Figure~\ref{fig:region}, characterizes the $(\alpha,\beta)$ feasibility region almost completely.
We establish this feasibility region by carefully analyzing the (random) de Bruijn graph formed by the $k$-mer set.
In this graph, the set of source sequences corresponds to a set of paths that cover all edges.
Our analysis shows that there are feasible $(\alpha,\beta)$ pairs where the $k$-mer de Bruijn graph is fairly complex, due to the presence of repeats across the source sequences, but nevertheless there is a unique way to decompose the graph into $m$ source paths.









\noindent {\bf Related work:}
An important line of related work deals with the \emph{coded} version of the problem we consider.
Since a sequence $\x$ containing no repeats of length $k-1$ can be reconstructed from the set of its $k$-mers \cite{ukkonen}, 
\cite{elishco2021repeat} studied how large a \emph{repeat-free code} can be.
They showed that, as long as $k = \lceil \beta \log n\rceil$, for $\beta > 1$, one can asymptotically build $(k-1)$-repeat-free codes with rate $1$. 
The setup where one seeks to reconstruct a codeword $\x$ from a noisy version of its set of $k$-mers have also received considerable attention~\cite{yehezkeally2021codes,marcovich2021reconstruction,chang2017rates,gabrys2018unique}.

A multi-sequence version of the coding problem in~\cite{elishco2021repeat} was  considered in \cite{yehezkeally2021codes}. Similar to our paper, they provide conditions on how $k$ needs to scale with $n$ and $m$ to guarantee that the asymptotic rate of all $k$-mer-reconstructible codes approaches 1.
Our work can be seen as an uncoded counterpart to \cite{yehezkeally2021codes}, where we characterize how $k$ needs to scale with $n$ and $m$ to allow the reconstruction of random source sequences with vanishing failure probability.
As an important distinction, we notice that the code constructions in \cite{yehezkeally2021codes} guarantee that there are no repeats across the $m$ sequences to be reconstructed, while in our setting, $m$ source sequences with repeats may still be reconstructible for some $(\alpha,\beta)$.

\section{Problem Setting and Main Results}

We consider $m$ source sequences $\x_1,\dots,\x_m$, each of which is generated as an i.i.d.~${\rm Ber}(\tfrac12)$ length-$n$ sequence.
The goal is to reconstruct the set of source sequences $\X := \{\x_1,\dots,\x_m\}$ from the set of all of their $(k+1)$-mers, i.e.,
\aln{
\Y(\X) := \{ \x_i[j:j+k] : 1\leq i \leq m,\, 1 \leq j \leq n - k \},
}
where $\x_i[a:b]$ is the substring of $\x_i$ between (and including) the $a$th and $b$th symbols. Here, we use $(k+1)$-mers rather than $k$-mers for notational convenience, and the final results are not affected. 
When clear from context, we write $\Y$ for $\Y(\X)$.
We will also use the notation $\x_i(a)$ to refer to the $a$th $k$-mer of $\x_i$, for $1\leq a \leq n-k+1$.

We consider an asymptotic regime where $m = n^\alpha$  and $k = \beta \log n$, for positive constants $\alpha$ and $\beta$. As it turns out, this is the asymptotic regime of interest, where the fundamental limits are non-trivial.
We say that $\Y$ reconstructs $\X$ if $\X$  is the unique set of $m$ length-$n$ sequences (up to relabeling) that could have generated $\Y$.
It is then natural to define a notion of feasibility as follows.

\begin{definition}
We say that $(\alpha,\beta)$ is a feasible pair if for $m = n^\alpha$ and $k = \beta \log n$, 
$\lim_{n \to \infty} \Pr( \Y \text{ reconstructs } \X) = 1.$
\end{definition}

The goal is to characterize the  feasibility region $\F \subset \R_+^2$, defined as the set of all feasible $(\alpha,\beta)$ pairs.

An equivalent and useful representation of the $k$-mer set $\Y$ is its de Bruijn graph $G(\Y)$.
The de Bruijn graph $G(\Y)$ can be obtained by taking the set of $k$-mers in $\Y$ (i.e., length-$k$ prefixes and suffixes of the $(k+1)$-mers in $\Y$) as the node set, and connecting two $k$-mers with a \emph{directed} edge if they are the prefix and suffix of a $(k+1)$-mer in $\Y$. 
Notice that the true set of sequences $\X$ can be seen as a set of paths on $G(\Y)$ that cover all the edges (edges can be used multiple times).

Our main result is to characterize, almost completely, the feasibility region $\F$ as follows:
\begin{theorem} \label{thm:main}
The following bounds on $\F$ hold:
\begin{itemize}
    \item $(\alpha,\beta) \in \F$ if $\beta > \max(2\alpha+1,\alpha+2)$,
    \item $(\alpha,\beta) \notin \F$ if $\beta < \max(2\alpha+1,\alpha+\tfrac32)$.
\end{itemize}
\end{theorem}

As illustrated in Figure~\ref{fig:region}, Theorem~\ref{thm:main} nearly completely characterizes $\F$.
In particular, for $\alpha > 1$, $(\alpha,\beta) \in \F$ if $\beta > 2\alpha + 1$, and $(\alpha,\beta) \notin \F$ if $\beta < 2\alpha + 1$.
For $\alpha < 1$ there is a small uncharacterized region of $(\alpha,\beta)$ pairs.

The rest of the paper is organized as follows. 
In Section~\ref{sec:norepeat}, as a warm-up, we characterize the repeat-free region, a (strict) subset of $\F$ where there are no length-$k$ repeats across the source sequences (and the source strings are thus reconstructible).
In Section~\ref{sec:achievability}, we prove the inner bound part of Theorem~\ref{thm:main}, and in Section~\ref{sec:converse}, we prove the outer bound part of Theorem~\ref{thm:main}.


\section{Repeat-Free Region}
\label{sec:norepeat}

We start by considering a simple scenario where $\mathcal{X}$ can be recovered correctly: when each $k$-mer is unique across all source sequences. 
In this case, it is well known that the de Bruijn graph $G(\Y)$ will contain $m$ disjoint paths, each of which corresponds to one of the $\x_i$s, and $\X$ is thus reconstructible.
We prove the following result.
\begin{lemma} \label{lem:norepeat}
If $\beta > 2\alpha + 2$, with probability tending to $1$ as $n\to \infty$, there are no repeated $k$-mers across or within $\x_i$s.
\end{lemma}
This region is marked in Figure~\ref{fig:region}.
Notice that it does not include all pairs of feasible $(\alpha,\beta)$ pairs.

To prove Lemma~\ref{lem:norepeat}, we note that, if $i \ne j$,
\aln{
\Pr\left(\x_i(a) = \x_j(b)\right) = 2^{-k} = 2^{-\beta \log n} = n^{-\beta}.
}
Hence, the probability that any two strings $\x_i$ and $\x_j$, for $i \ne j$ have at least one $k$-mer in common is upper-bounded using the union bound by
\begin{align*}
\binom{m}{2} n^2 n^{-\beta} \leq m^2 n^{2 - \beta} \leq n^{2\alpha + 2 -\beta}.
\end{align*}

Additionally, we need to consider the case where a single string contains a repeat $k$-mer.
In this case, not all of the $k$-mers are independent due to potential overlaps.
However, a careful analysis reveals that, even if $|b-a| < k$ and $\x_i(a)$ and $\x_i(b)$ overlap, we still have $\Pr\left(\x_i(a) = \x_i(b)\right) = n^{-\beta}$ (see Section~\ref{app:prB} for a similar analysis).
We conclude that the probability that there is repeated $k$-mer within some $\x_i$ is at most $m n^2 n^{-\beta} = n^{\alpha+2-\beta}$, and the probability that any $k$-mer is repeated across or within $\x_1,\dots,\x_m$ is upper-bounded by
\begin{align*}
n^{2\alpha+2-\beta} + n^{\alpha+2-\beta},
\end{align*}
which tends to zero for $\beta>2\alpha+2$. 
A converse result can also be shown; i.e., for $\beta < 2\alpha + 2$, there will be repeats with probability tending to $1$ (but we do not describe that).
In the next section, we show that there are feasible $(\alpha,\beta)$ pairs that are in this repeat-abundant regime.

\section{Inner Bound to Feasibility Region}
\label{sec:achievability}

In order to characterize $\F$, we define the error event $\E = \{ \Y \text{ does not reconstruct } \X\}$, and seek to characterize the asymptotic behavior of $\Pr(\E)$ for a given choice of $(\alpha,\beta)$.
To do so, we will need to define several  events related to repeats: 
\aln{
\A = \{ & \exists i,a,b:\x_i(a)=\x_i(b) \} \\
\B = \{ & \exists i,j,l,a,b,c,d:  \x_i(a)=\x_j(c),\x_i(b)=\x_l(d), \\ 
&  0\leq b-a < k, \text{ and either } 
\{ j \ne l \} \text{ or } \\
& \{ j = l \text{ and } d-c\neq b-a \}
\} \\
\C = \{ & \exists i,j,a:\x_i(1)=\x_j(a)\} 
\cup \{\exists i,j,a:\x_i(n')=\x_j(a)\}
\\
\D = \{ & \exists i,j,a: \x_i(a) = \x_j(a)
\}
}
where we let $n' = n-k+1$ for conciseness.





    

    
Event $\A$ is the event that there is at least one intra-sequence repeat.
Event $\B$ is the event that there are two pairs of repeats with an overlap (i.e., $\x_i(a)$ and $\x_i(b)$ overlap in $\x_i$), but they are not simply a longer length-$t$ repeat for $t > k$.
Notice that $\B$ includes the event that a triple repeat $\x_i(a) = \x_j(b) = \x_l(c)$ occurs.
Event $\C$ is the event that the first or last $k$-mer of some $\x_i$ appears somewhere else. 
Event $\D$ is the event that the $a$th $k$-mer of two sequences $\x_i$ and $\x_j$ is the same.
Notice that events $\A$, $\B$ and $\C$ are not error events per se (they each do not imply $\E$), but they will be useful in the analysis.
On the other hand, $\D \Rightarrow \E$, since given $\D$ it is possible to replace $\x_i$ and $\x_j$ with two incorrect sequences
\aln{
& \tilde \x_i = \x_i[1:a-1]\x_j[a:n] \\
& \tilde \x_j = \x_j[1:a-1]\x_i[a:n].
}
We will upper bound the error probability as 
\al{
\Pr(\E) & \leq \Pr(\A \cup \B \cup \C \cup \D \cup \E) \nonumber\\
& \leq \Pr(\A) + \Pr(\B) + \Pr(\C) + \Pr(\D) + \nonumber \\
& \quad + \Pr(\E \cap \bar \A \cap \bar \B \cap \bar \C \cap \bar \D ) \label{eq:error_broken_into_events}
}
and analyze the conditions for each term to go to zero.
First we notice that, by the union bound, 
\al{
\Pr(\A)\leq mn^2 n^{-\beta} = 
n^{\alpha + 2 -\beta},
}
from which we see that $\Pr(\A) \to 0$ if $\beta > \alpha + 2$.
Similarly, 
\al{
& \Pr(\B) \leq m^3n^3 2k n^{-2\beta} + m^2 n^2 (2k)^2 n^{-2\beta}  \nonumber \\
& \quad \quad \quad \leq  n^{3\alpha + 3 - 2\beta + o(1)},
\label{eq:prB}
\\
& \Pr(\C)  \leq 2m^2n n^{-\beta}= 2n^{2\alpha+1-\beta}, \\
& \Pr(\D)  \leq m^2 n n^{-\beta} = 4n^{2\alpha +1 -\beta}.
}
We point out that to establish (\ref{eq:prB}), we need a careful analysis of the case where $\x_j(c)$ and $\x_j(d)$ also overlap, presented in Lemma~\ref{lm:independent}.

As a result we see that, if $\beta > \max(\alpha+2,\tfrac32\alpha+\tfrac32,2\alpha+1)$, $\Pr(\A) + \Pr(\B) + \Pr(\C) + \Pr(\D) \to 0$.
What remains is to bound $ \Pr(\E \cap  \bar \A \cap \bar \B \cap \bar \C \cap \bar \D)$.

Now suppose $\bar \A \cap \bar \B \cap \bar \C \cap \bar \D$ holds. 
Let the multiplicity $\mu_{\X}(\bfv)$ of an node $\bfv$ in $G(\Y)$ be the number of times it is traversed by paths in $\X$.
Notice that $\bar \B$ guarantees that no node can be traversed three or more times, so given $\bar \B$, node multiplicities can only be $1$ or $2$. 

The proof of the following lemma is in Appendix~\ref{app:lemmas}.

\begin{lemma}\label{lm:fulldet}
    Suppose $\bar\A$, $\bar\B$, and $\bar\C$ hold. Then the multiplicity of every node in $G(\Y)$ is fully determined. In other words, every valid solution 
    of $G(\Y)$ traverses a given node in $G(\Y)$ the same number of times.  
\end{lemma}


Now suppose $\E$ occurs. 
By definition, there is at least one $\tilde \X \ne \X$ such that $\Y(\tilde \X) = \Y(\X)$. 
Let $\tilde \X$ be one such choice with minimum set difference $|\X - \tilde\X|$.
Let $|\X - \tilde\X| = c$. By Lemma \ref{lm:fulldet}, we know that for all nodes $\bfv$,
\al{\mu_{\X}(\bfv)=\mu_{\tilde\X}(\bfv):=\mu_{G}(\bfv).}
Additionally, 
\al{
\mu_{\X\cap\tilde{\X}}(\bfv) + \mu_{\X-\tilde{\X}}(\bfv) = \mu_{\X}(\bfv) &= \mu_G(\bfv)\\
\mu_{\X\cap\tilde{\X}}(\bfv) + \mu_{\tilde{\X}-\X}(\bfv) = \mu_{\tilde\X}(\bfv) &= \mu_G(\bfv),
}
implying that $\mu_{\X-\tilde{\X}}(\bfv) = \mu_{\tilde{\X}-\X}(\bfv)$.

We now build a ``difference de Bruijn graph'' $\tilde G(\Y,\X,\tilde\X)$ on the $(k+1)$-mers in $\Y$ that are present in $\X-\tilde\X$. It is easy to see that $\tilde G(\Y,\X,\tilde\X)$ is the same as the subgraph of $G(\Y)$ induced by the edges corresponding to $(k+1)$-mers in $\X-\tilde\X$ with nodes $\{\bfv:\mu_{\X-\tilde\X}(\bfv)>0\}$. Note that the nodes and their multiplicities are equivalent to those of the set $\{\bfv:\mu_{\tilde\X-\X}(\bfv)>0\}$, and the set of edges in $\tilde G(\Y,\X,\tilde\X)$ must also correspond to all of the $(k+1)$-mers in $\Y$ that are present in $\tilde\X-\X$.  
Therefore $\tilde G(\Y,\X,\tilde\X)$ is the same as the de Bruijn graph on the $(k+1)$-mers in $\Y$ that are present in $\tilde\X-\X$; i.e., $\tilde G(\Y,\X,\tilde\X) = \tilde G(\Y,\tilde\X,\X)$. 
Notice that $\tilde G (\Y,\X,\tilde \X)$ is a de Bruijn graph for both the set of $c$ missing true paths $\X - \tilde \X = \{\x_1,\dots,\x_c\}$ and for the set of incorrectly reconstructed paths $\tilde \X - \X = \{\tilde \x_1,\dots,\tilde \x_c\}$.

Define a \textit{maximal shared subpath} in $\tilde G(\Y,\X,\tilde\X)$ as a directed sequence $\mathbf{p}$ of nodes with maximal length such that $\mu_{\X-\tilde\X}(\bfv)=2$ for all nodes $\bfv\in\mathbf{p}$. We have the following lemma.


\begin{lemma}\label{lm:cpaths}
    Given $\bar \A,\bar \B, \bar \C, \bar \D$,
    there are at least $c$ maximal shared subpaths in $\tilde G(\Y,\X,\tilde\X)$.
\end{lemma}

\begin{IEEEproof}
Given $\bar \A$, $\bar \B$, $\bar \C$, and Lemma~\ref{lm:fulldet}, the multiplicities of all nodes in $\tilde G(\Y,\X,\tilde\X)$ are fully determined and must be either $1$ or $2$.
Now consider the paths corresponding to $\tilde \x_1,\dots,\tilde \x_c$ in $\tilde G(\Y,\X,\tilde\X)$.
We claim that, given $\bar \C$ and $\bar \D$, each $\tilde \x_i$ must traverse at least two maximal shared subpaths.
Otherwise, suppose $\tilde \x_i$ only traverses a single maximal shared subpath,
say $\x_1[s:t]$, which is shared between $\x_1$ and $\x_2$.
Given $\bar \C$, $\tilde \x_i$ must start and end at nodes with zero in-degree and out-degree respectively.
Hence we must be able to write $\tilde \x_i$, without loss of generality, as
\aln{
\tilde \x_i = \x_1[1:s-1]\x_{1}[s:t]\x_2[t+1:n]
}
But this implies that $\x_1(s) = \x_2(s)$, which contradicts $\bar \D$.

Finally, since each $\tilde \x_i$, $i=1,\dots,c$, must traverse at least two maximal shared subpaths and, given $\bar \B$, every maximal shared subpath can be shared by at most two true paths $\x_i$, we conclude that there must be at least $c$ maximal shared subpaths in $\tilde G(\Y,\X,\tilde\X)$.
\end{IEEEproof} 

\begin{figure}
    \centering
    \includegraphics[width=0.9\columnwidth]{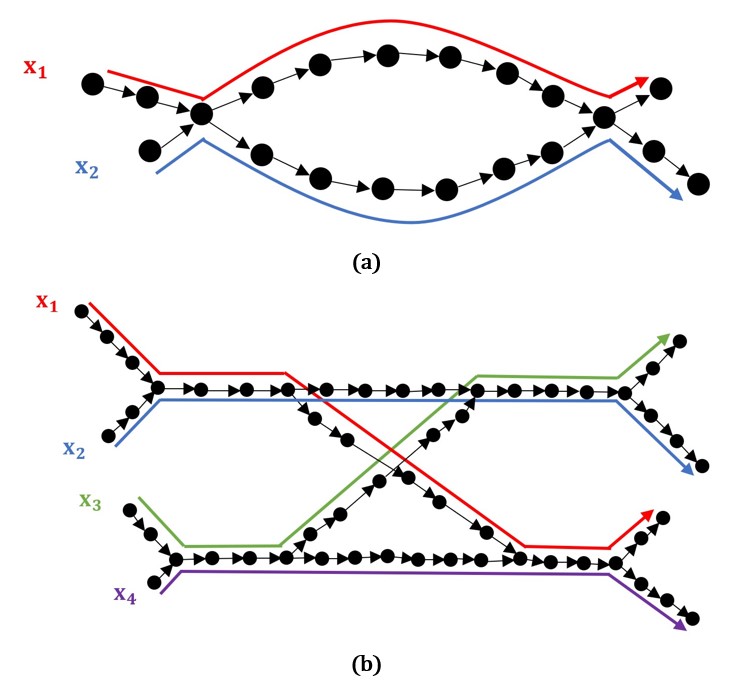}
    \caption{Two examples of difference de Bruijn graphs $\tilde G$. (a) Example $\tilde{G}$ for $c=2$. The red and blue paths represent the correct strings, while the other reconstruction follows the paths red--blue--red and blue--red--blue. (b) Example $\tilde{G}$ for $c=4$. The red, blue, green, and purple paths represent the correct strings, while a possible incorrect reconstruction follows the paths red--blue--green, blue--red--purple, green--purple--red, purple--green--blue.}
    \label{fig:examples}
\end{figure}

Figure~\ref{fig:examples} shows examples of the graph $\tilde G(\Y,\X,\tilde\X)$ given $\bar \A, \bar \B,\bar \C, \bar \D$, for $c = 2$ and $c=4$.
Notice that we have $c$ paths and $c$ maximal shared subpaths.
Also notice that, in both examples, there exist two choices for the set of $c$ paths.

Lemma~\ref{lm:cpaths} allows us to analyze $\Pr(\E)$ by partitioning it into events $\E_c := \E \cap \{ |\tilde \X - \X| = c \}$.
Notice that a maximal shared subpath between $\x_i$ and $\x_j$ occurs with probability at most $n^{-\beta}$.
Moreover, by the minimality of $\tilde \X$, each $\tilde \x_i$ must contain at least one maximal shared subpath.
By the union bound, we conclude that 
\aln{
\Pr( & \E_c \cap \bar \A \cap \bar \B \cap \bar \C \cap \bar \D) \\
& \leq \binom{m}{c} n^c (cn)^c n^{-\beta c}
\leq m^c n^{2c} n^{-\beta c} = n^{c(\alpha + 2 - \beta)},
}
where we used the fact that $\tbinom{m}{c} \leq (m/c)^c$.
Summing over all values of $c$, we obtain
\aln{
\Pr(\E \cap \bar \A \cap \bar \B \cap \bar \C \cap \bar \D) & \leq \sum_{c=2}^{m} n^{c(\alpha + 2 - \beta)} \\
& = 
\frac{n^{2(\alpha + 2 - \beta)} (1  - n^{(m-1)(\alpha + 2 - \beta )}) }{1- n^{\alpha + 2 - \beta }},
}
which tends to zero as $n \to \infty$ provided that $\beta > \alpha + 2$.
Plugging this back into \eqref{eq:error_broken_into_events}, we conclude that any $(\alpha,\beta)$ pair with 
$\beta > \max(\alpha+2,\tfrac32\alpha+\tfrac32,2\alpha+1) = \max(\alpha+2,2\alpha+1)$ is feasible.
In other words, we showed that 
\aln{
\{ (\alpha,\beta) : \beta > \max( \alpha + 2, 2\alpha + 1)\} \subseteq \F,
}
concluding the proof of the achievability part of Theorem~\ref{thm:main}.

\section{Outer Bound to Feasibility Region}
\label{sec:converse}

In this section, we will prove the outer bound part of Theorem~\ref{thm:main}.
First, we will show that $\Pr(\E) \not\to 0$ as $n \to \infty$, if $\beta < 2\alpha + 1$.
Second, we will show that $\Pr(\E) \not\to 0$ if $\beta < \alpha + \tfrac32$.
To do so, we will show that in  those regimes, with positive probability, there will be multiple reconstructions.


Define $Z_{i,j,a}=\mathbbm{1}_{\x_i(a)=\x_j(a)}$ for $i<j$; i.e., $Z_{i,j,a}$ indicates whether $\x_i$ and $\x_j$ share a $k$-mer at a common position $a$. 
Let $V=\sum_{i=1}^{m-1}\sum_{j=i+1}^m\sum_{a=1}^{n-k+1}Z_{i,j,a}$, and note that $\D=\{V>0\}$. 
In particular, since any $Z_{i,j,a} = 1$ implies that there are multiple valid reconstructions, we have that $\Pr(\E) \geq \Pr(V > 0)$.
By the Paley-Zygmund inequality, 
we then have
\al{
\Pr(\E) \geq \Pr(V>0)\geq \frac{\mathbb{E}[V]^2}{\mathbb{E}[V^2]}.
}
It is straightforward to see $\mathbb{E}[Z_{i,j,a}]=n^{-\beta}$, and by the linearity of expectation, 
\begin{equation*}\label{eq:numerator}
\mathbb{E}[V]=\binom{m}{2}(n-k+1)n^{-\beta}.
\end{equation*}
The computation of $E[V^2]$ requires more care in handling $\Pr(\x_i(a)=\x_j(a),\x_s(b)=\x_t(b))$. First, we note that when the pairs $(i,j)$ and $(s,t)$ are disjoint, the events $\x_i(a)=\x_j(a)$ and $\x_s(b)=\x_t(b)$ are independent:
\aln{
\Pr(\x_i(a)=\x_j(a),\x_s(b)=\x_t(b))&=n^{-2\beta}.
}
The events are also independent when $i=s$ but $j\neq t$, or $j=t$ but $i\neq s$,
\aln{
\Pr(\x_i(a)=\x_j(a),\x_i(b)=\x_t(b))=n^{-2\beta},
}
and when $i=s$, $j=t$, and $|a-b|\geq k$. However, when $i=s$, $j=t$, and $|a-b|< k$, 
\aln{
\Pr(\x_i(a)=\x_j(a),\x_i(b)=\x_j(b))
= 2^{-(k + |a-b|)}
\leq n^{-\beta}
}
as there are $|a-b|$ symbols of overlap between $\x_i(a)$ and $\x_j(b)$. Therefore,
\begin{align}
    \mathbb{E}[V^2]&=\mathbb{E}\left[ \sum_{i=1}^{m-1}\sum_{j=i+1}^m\sum_{a=1}^{n-k+1} Z_{i,j,a} \sum_{s=1}^{m-1}\sum_{t=s+1}^m\sum_{b=1}^{n-k+1} Z_{s,t,b} \right]\nonumber\\
    &\leq m^4n^{2-2\beta} + 2m^2n^{1-\beta}k.\label{eq:denominator}
\end{align}
Combining $\mathbb{E}[V]$ and (\ref{eq:denominator}), we have that
\al{
\Pr(\E)&\geq \frac{\binom{m}{2}^2(n-k+1)^2n^{-2\beta}}{m^4n^{2-2\beta}+2m^2n^{1-\beta}k}\\
& \sim \frac{m^4 n^2 n^{-2\beta}}{4m^4n^{2-2\beta}+8m^2n^{1-\beta}k} \label{eq:asym} \\
& = \frac{1}{4+\frac{8\beta\log n}{n^{2\alpha+1-\beta}}} \label{eq:asym2}
}
where (\ref{eq:asym}) is asymptotically equivalent as $n\to\infty$. 
Finally, we notice that if  $\beta<2\alpha+1$, (\ref{eq:asym2}) converges to $\frac{1}{4}$, implying that $\Pr(\E)$ is bounded away from zero and $(\alpha,\beta) \notin \F$. 

Next, we show that $\Pr(\E)\not\to 0$ when $\beta<\alpha+\frac{3}{2}$. 
For this regime, 
we will define the event $\cH$ that there are two strings $\x_i$, $\x_j$ and indices $a$, $b$, $c$ such that 
$$
\x_i(a)=\x_j(c),\x_i(b)=\x_j(c+b-a).
$$
This is illustrated in Figure~\ref{fig:examples}(a).
Essentially, we have two repeated $k$-mers across sequences $\x_i$ and $\x_j$, and the gap between them is the same in both sequences.
Given $\cH$, an incorrect reconstruction exists because we can swap the middle parts of $\x_i$ and $\x_j$, creating
\aln{
& \tilde \x_i = \x_i[1:a-1]\x_j[c:c+b-a-1]\x_i[b:n] \\
& \tilde \x_j = \x_j[1:c-1]\x_i[a:b-1]\x_j[c+b-a:n],
}
which can be verified to be two length-$n$ sequences. 
We will show that $\Pr(\cH) \not\to 0$ as $n\to\infty$. 

Define $W_{i,j,\mathbf{a}}=\mathbbm{1}_{\x_i(a_1)=\x_j(a_3),\x_i(a_2)=\x_j(a_3+a_2-a_1)}$ for $\mathbf{a}=(a_1,a_2,a_3)$ and let $U=\sum_{i,j}\sum_{\mathbf{a}}W_{i,j,\mathbf{a}}$. 
We can again use the Paley-Zygmund inequality to find a lower bound on $\Pr(U>0)$. The second moment $\mathbb{E}[U^2]$ can be computed similarly to $\mathbb{E}[V^2]$, 
considering separately the terms of the form $\mathbb{E}[W_{ij\mathbf{a}}W_{ij\mathbf{b}}]$
where the $k$-mers indicated by $\mathbf{a}$ overlap with those indicated by $\mathbf{b}$:
\al{
\mathbb{E}[U^2]&=\mathbb{E}\left[ \sum_{i=1}^{m-1}\sum_{j=i+1}^m\sum_{\mathbf{a}}W_{i,j,\mathbf{a}} \sum_{s=1}^{m-1}\sum_{t=s+1}^{m}\sum_{\mathbf{b}}W_{s,t,\mathbf{b}} \right]\nonumber\\
&\leq (m^2n^{3}n^{-2\beta})^2 + m^2n^{3}(2k)^3n^{-2\beta}\label{eq:u2}
}
where the first term in (\ref{eq:u2}) covers all $k$-mers that are independent of each other and the second term covers the case where $i=s$, $j=t$, and $a_i$ overlaps with $b_i$ for $i=1,2,3$. 

To compute the first moment, $\mathbb{E}[U]$, we must account for which assignments of $\mathbf{a}$ are possible.
More precisely, we need to count the vectors $(a_1,a_2,a_3) \in [1:n']^3$ such that $a_1 < a_2$ and $a_2+a_3-a_1 \in [1:n']$, where $n' = n - k +1$.
This is cumbersome, so we consider a lower bound by considering $a_1 \in [1:n'/4]$, $a_2 \in [n'/4+1:n'/2]$, and $a_3 \in [n'/4+1:n'/2]$.
Notice that for any of the $(n'/4)^3$ choices, $a_2 + a_3 - a_1 \in [1:n']$.
We obtain the (rather loose) lower bound
\al{
\mathbb{E}[U] \geq \binom{m}{2} \left( \frac{n'}{4}\right)^3 n^{-2\beta}.
}
Therefore, we can lower bound the error probability as
\aln{
\Pr(\E) & \geq \frac{\mathbb{E}[U]^2}{\mathbb{E}[U^2]} 
\geq \frac{\binom{m}{2}^2 (n'/4)^6 n^{-4\beta}}{m^4n^{6}n^{-6\beta} + m^2n^{3}(2k)^3n^{-2\beta}}
\\
& \sim  \frac{4^{-7} m^4 n^{6-4\beta}}{m^4 n^{6-4\beta} + 8 m^2 k^3 n^{3-2\beta}} 
\\
& = \frac{4^{-7}}{1 + \frac{8 k^3}{n^{2^{\alpha + 3-2\beta}}},
}
}
which tends to $4^{-7}$ as $n \to \infty$, as long as $\beta < \alpha + \tfrac32$.
We conclude that for any $(\alpha,\beta)$ such that $\beta < \alpha + \tfrac32$ or $\beta < 2\alpha + 1$, $\Pr(\E) \not\to 0$ as $n \to \infty$.
This concludes the proof of the outer bound in Theorem~\ref{thm:main}.


\section{Concluding Remarks}

In this paper, we have characterized the feasibility of a multiple sequence reconstruction problem using $k$-mers for a large region of $(\alpha,\beta)$ parameters. We identified a region where the $k$-mers of each string are disjoint with high probability and correct recovery of the source sequences is trivial, as well as a region where, while repeats exist, the only possible reconstruction of the source sequences is the correct one. In the latter repeat-abundant region, careful examination of the de Bruijn graph $G(\Y)$ reveals the unique set of paths covering the graph. Finally, we have determined a region where incorrect reconstructions exist with positive probability. 

The above findings have characterized nearly all $(\alpha,\beta)$ pairs for $\alpha>0$, $\beta>0$, except for a small region between $\beta=\max(2\alpha+1,\alpha+\frac{3}{2})$ and $\beta=\alpha+2$. We believe that part of the remaining region may be where the solution is not unique, and close analysis of additional difference de Bruijn graph topologies for a larger number of strings and shared $k$-mers might result in a larger red region in Figure \ref{fig:region}.

This work invites future study in several directions. One of these is generalization to other source distributions; some results here may be immediately trivially extended to a general i.i.d. source. Additionally, motivated by real-world DNA sequencing technologies, future work may analyze source reconstruction from noisy $k$-mers or a random number of copies of each $k$-mer. 

\section*{Acknowledgements}

The work of I.S. was supported in part by the National Science Foundation under CCF grants 2007597 and 2046991.
The authors would like to thank the anonymous reviewers of this manuscript for their detailed feedback and suggestions for improvement. 

{\footnotesize
\bibliographystyle{IEEEtran}
\bibliography{references.bib}
}

\newpage

\appendices 

\section{Proof of Equation (\ref{eq:prB})}
\label{app:prB}

Taking the union over the events $\{j\neq l\}$ and $\{j=l\}$, the probability of $\B$ can be bounded as
\al{
\Pr(\B)&\leq m^3n^32kn^{-2\beta} + m^2n^2(2k)^2P_{i,j}(a,b,c,d)
}
where $P_{i,j}(a,b,c,d)=\Pr(\x_i(a)=\x_j(c),\x_i(b)=\x_j(d))$ is shown in the following lemma to be upper-bounded by $n^{-2\beta}$ when $d-c\neq b-a$.
\begin{lemma}\label{lm:independent}
For any two sequences $\x_i$, $\x_j$ and indices $a$, $b$, $c$, $d$, with $b>a$ and  either $b-a>1$ or $|d-c|>1$, the probability that $\x_i(a)=\x_j(c)$ and $\x_i(b)=\x_j(d)$ is upper-bounded by $n^{-2\beta}$.
\end{lemma}
\begin{IEEEproof}
If either $b-a\geq k$ or $|d-c|\geq k$, then it is clear that either $\x_i(a)$ and $\x_i(b)$ or $\x_j(c)$ and $\x_j(d)$, respectively, are independent of each other. Therefore, $\Pr(\x_i(a)=\x_j(c),\x_i(b)=\x_j(d))=\Pr(\x_i(a)=\x_j(c))\Pr(\x_i(b)=\x_j(d))=n^{-2\beta}$. 

If both $b-a<k$ and $|d-c|<k$, then closer analysis is required, since $\x_i(a)$ overlaps with $\x_i(b)$ and $\x_j(c)$ overlaps with $\x_j(d)$ in this case. 
Note that if $b-a=|d-c|<k$, $\x_i(a)=\x_j(c)$, and $x_i(a+1)\neq x_j(c+1)$, then $\Pr(\x_i(b)=\x_j(d))=0$, since $b$ and $d$ will contain the differing symbols in $x_i(a+1)$ and $x_j(c+1)$, respectively. 

We first assume that both $a<b$ and $c<d$. Let $t_i=k-(b-a)$ and $t_j=k-(d-c)$; without loss of generality, we assume that $t_i>t_j$ (if not, we can swap the labels of $x_i$ and $x_j$). These values represent the overlap in symbols between the substrings of each string. If $\x_i(a)=\x_j(c)$, then 
\begin{align*}
    x_i(a)&=x_j(c)\\
    &\dots\\
    x_i(a+k-1)&=x_j(c+k-1),
\end{align*}
and if $\x_i(b)=\x_j(d)$, then
\begin{align*}
    x_i(b)&=x_j(d)\\
    &\dots\\
    x_i(b+k-1)&=x_j(d+k-1).
\end{align*}
Furthermore, if $\x_i(a)$ and $\x_i(b)$ overlap by $t_i$ symbols, 
\begin{align*}
    x_i(a+k-t_i)&=x_i(b)\\
    &\dots\\
    x_i(a+k-1)&=x_i(b+t_i-1).
\end{align*}
 Similarly, if $\x_j(c)$ and $\x_j(d)$ overlap by $t_j$ symbols,
\begin{align*}
    x_j(c+k-t_j)&=x_j(d)\\
    &\dots\\
    x_j(c+k-1)&=x_j(d+t_j-1).
\end{align*}
In other words, the last $t_i$ symbols of $\x_i(a)$ are equal to the first $t_i$ symbols of $\x_i(b)$, and the last $t_j$ symbols of $\x_j(c)$ are equal to the first $t_j$ symbols of $\x_j(d)$. Since $\x_i(a)=\x_j(c)$ and $\x_i(b)=\x_j(d)$, this implies that the last $t_j$ symbols of $\x_i(a)$ are equal to the first $t_j$ symbols of $\x_i(b)$, or 
\begin{align*}
    \x_i(a+k-t_j)&=\x_i(b)\\
    &\dots\\
    \x_i(a+k-1)&=\x_i(b+t_j-1).
\end{align*}
The probability that there are these $t_j$ replicated symbols in $\x_i(a)$ is $2^{-t_j}$, the probability that $\x_i(a)=\x_j(c)$ is $2^{-k}$, 
and the probability that $\x_i(b)=\x_j(d)$ given $\x_i(a)=\x_j(c)$ and the requisite repetition of symbols in $\x_i(a)$ is $2^{-(k-t_j)}$. 
Combining these probabilities, 
\begin{align*}
    \Pr(\x_i(a)=\x_j(c),\x_i(b)=\x_j(d))=n^{-2\beta}.
\end{align*}
The case where $a<b$ and $c>d$ can be shown similarly; here, the last $t_j$ symbols of $\x_i(b)$ must be equal to the first $t_j$ symbols of $\x_i(a)$. 
\end{IEEEproof}

Therefore,
\al{
\Pr(\B)&\leq m^3n^32kn^{-2\beta}\\
&=n^{3\alpha+3-2\beta + \frac{\log(\beta\log n)}{\log n}}\\
&=n^{3\alpha+3-2\beta+o(1)}.
}
\section{Proof of Lemma~\ref{lm:fulldet}}
\label{app:lemmas}

\begin{IEEEproof}[Proof of Lemma~\ref{lm:fulldet}] This can be proven by demonstrating an algorithm that can sequentially and unambiguously determine the multiplicity of each node.  
According to $\bar\C$, the first (start) and last (end) nodes of each string are known and have multiplicity 1. The node-labeling algorithm can thus label the $m$ start nodes with 1 and continue towards the end nodes until a node that has two incoming edges is reached. 
Due to $\bar\B$, this node can be shared by a maximum of two sequences, and due to $\bar\A$, each of those strings can contain the node's corresponding $k$-mer exactly once, so the multiplicity of this node must be 2. This node is then followed either by an edge pointing to a node that also has multiplicity 2 or two edges pointing to nodes of multiplicity 1. 
\end{IEEEproof}

\end{document}